\documentclass[12pt]{iopart}
\usepackage{iopams}
\usepackage{graphicx}
\usepackage{graphics}

\begin{document}

\newcommand{\vc}[1]{\mathbf{#1}}

\title{Nonlocal electron-phonon interaction as a source of dynamic charge stripes in the cuprates}
\author{Claus Falter, Thomas Bauer}
\address{Institut f\"ur Festk\"orpertheorie, Westf\"alische Wilhelms-Universit\"at,\\
Wilhelm-Klemm-Str.~10, 48149 M\"unster, Germany}
\date{\today}

\begin{abstract}
We calculate for La$_{2}$CuO$_{4}$ the phonon-induced
redistribution of the electronic charge density in the insulating,
the underdoped pseudogap and the more conventional metallic state
as obtained for optimal and overdoping, respectively. The
investigation is performed for the anomalous
high-frequency-oxygen-bond stretching modes (OBSM) which
experimentally are known to display a strong softening of the
frequencies upon doping in the cuprates. This most likely generic
anomalous behaviour of the OBSM has been shown to be due to a
strong nonlocal electron-phonon interaction (EPI) mediated by
charge fluctuations on the ions. The modeling of the competing
electronic states of the cuprates is achieved in terms of
consecutive orbital selective incompressibility-compressibility
transitions for the charge response. We demonstrate that the
softening of the OBSM in these states is due to nonlocally induced
dynamic charge inhomogenities in form of charge stripes along the
CuO bonds with different orbital character. Thus, a multi-orbital
approach is essential for the CuO plane. The dynamic charge
inhomogeneities may in turn be considered as precursors of static
charge stripe order as recently observed in
La$_{2-x}$Ba$_{x}$CuO$_{4}$ in a broad range of doping around
$x\,=\,1/8$. The latter may trigger a reconstruction of the Fermi
surface into small pockets with reduced doping. We argue that the
incompressibility of the Cu$3d$ orbital and simultaneously the
compressibility of the O$2p$ orbital in the pseudogap state seems
to be required to nucleate dynamic stripes.
\end{abstract}


\maketitle

In recent work \cite{Falter06, Bauer08} we have developed a
description of the electronic charge response and phonon dynamics
for $p$-doped and $n$-doped cuprate high-temperature
superconductors (HTSC's). This approach is based upon a
microscopic modeling of the change of the charge response across
the insulator-metal transition in terms of the compressibility of
the electronic system as a primary tool to characterize the
corresponding competing ground states depending on doping. This is
achieved starting from the insulating state via a metallic
pseudogap state to a more conventional metallic state by
consecutive orbital selective incompressibility-compressibility
transitions in terms of strict sum rules of the charge response
given in Ref. \cite{Falter06} and listed below. These sum rules
are a proper instrument to describe the competition between
electronic localisation and itinerancy which is at the heart of
the physics in the cuprates.

In the conventional metallic state the electronic partial density of
states (PDOS) at the Fermi level $Z_{\kappa}\,(\varepsilon_{F})$ is
related to the polarizability matrix in the longwavelength limit
$(\vc{q}\,\to\,\vc{0})$ according to
\begin{equation} \label{1}
\sum\limits_{\kappa'}\,\Pi_{\kappa\kappa'}\,(\vc{q}\,\to\,\vc{0})\,=\,Z_{\kappa}\,(\varepsilon_{F})\,\,,
\end{equation}
and the total density of single particle states at energy
$\varepsilon$ is given by
\begin{equation} \label{2}
Z\,(\varepsilon)\,=\,\sum\limits_{\kappa}\,Z_{\kappa}\,(\varepsilon)\,\,.
\end{equation}
$\kappa$, $\kappa'$ are the orbital degrees of freedom in the
elementary cell of the crystal. In our calculation for LaCuO, Cu$3d$,
Cu$4s$ and O$2p$ orbitals in the CuO plane are taken into account.
$\vc{q}$ is a wavevector from the first Brillouin zone (BZ). On the
other hand, for the insulating state we obtain the sum rules
\begin{equation} \label{3}
\sum\limits_{\kappa'}\,\Pi_{\kappa\kappa'}\,(\vc{q}\,\to\,\vc{0})\,=\,O\,(q)
\end{equation}
and
\begin{equation} \label{4}
\sum\limits_{\kappa,\,\kappa'}\,\Pi_{\kappa\kappa'}\,(\vc{q}\,\to\,\vc{0})\,=\,O\,(q^{2})\,\,.
\end{equation}
The sum
$\sum\limits_{\kappa,\,\kappa'}\,\Pi_{\kappa\kappa'}\,(\vc{q}\,\to\,\vc{0})$
is equal to $\rho^{2}\,K$ with $\rho$ the average density and $K$ the
compressibility of the electronic system. The latter provides a measure
of the charge gap in the electronic spectrum because $K$ vanishes as a
function of the chemical potential in the gap region. Moreover, $1/K$
provides a measure of the change of the chemical potential with
particle number.

Approaching the delocalization-localization transition from the
conventional metallic region representative for optimal and
overdoping, when $p$-doping is decreased the Cu$3d$ component of
the wave function is admitted to become incompressible,
insulator-like in the underdoped state according to our modeling.
This means that the correlated 3d electrons are allowed to become
more localized as compared with the conventional metallic state
and disappear from the Fermi surface (FS). In terms of the sum
rule we then have
\begin{equation} \label{5}
\sum\limits_{\kappa'}\,\Pi_{\kappa\kappa'}\,(\vc{q}\,\to\,\vc{0})\,=\,\left\{
\begin{array}{cll}
O\,(q) & , & {\rm Cu}3d \\
Z_{\kappa}\,(\varepsilon_{F}) & , & {\rm O}2p\,\,.
\end{array} \right.
\end{equation}
Accordingly, the partial density of states is suppressed at the Fermi
level for the correlated, localized Cu$3d$ orbitals (orbital intrinsic
incompressibility due to correlation) but not for the more delocalized
O$2p$ orbitals at the O$_{xy}$ sublattices in the CuO plane, where the
holes are predominantly injected in the $p$-type cuprates. So, we have
an orbital selective compressible, metallic charge response only for
the O$2p$ states with a renormalized PDOS
$Z_{\kappa}\,(\varepsilon_{F})$ and a loss in the total density of
state at $\varepsilon_{F}$ ({\em pseudogap}) because of the
incompressibility of the Cu$3d$ states according to equation (\ref{5}).
The loss in spectral weight also is never recovered on entering the
superconducting state. The incompressible regions in the pseudogap
state compete with overall metallic behaviour and with
superconductivity.

Our modeling of the pseudogap state in terms of compressible,
metallic O$2p$ states and Cu$3d$ states treated as incompressible
and disappearing from the FS because of strong correlation is
supported by calculations performed within the
self-interaction-corrected (SIC) local spin density (LSD)
approximation for La$_{2}$CuO$_{4}$ \cite{Svane92}. Here it is
shown that the self-interaction pulls the occupied localized
Cu$3d$ states below the O$2p$ band, only little Cu$3d$ admixture
is found in the O$2p$ band.

Furthermore, our modeling of the pseudogap state corresponds with a
smaller Fermi volume as compared with the conventional metallic
Fermi-liquid state because the Fermi volume is determined by the
density of the compressible metallic charge carriers (holes) of
dominantly O$2p$-type alone while the Cu3d states are incompressible as
in the Mott insulator and cannot contribute to the Fermi volume quite
in contrast to the conventional metallic state where both the O$2p$ and
the Cu$3d$ states are compressible metallic leading to a large Fermi
volume. The pseudogap state is separated from the conventional overall
compressible Fermi-liquid state by an orbital selective
compressibility-incompressibility transition of the Cu$3d$ orbital. In
the context of our modeling of the pseudogap state it is interesting to
point to another modeling of this state within the framework of the so
called fractionalized Fermi liquid. For a recent discussion see e.g.
Ref. \cite{Vojta} and references therein. In this approach an idea is
important which is based on the concept of an orbital-selective Mott
transition. For the latter, e.g. in a two band model, we have one
metallic band with Fermi-liquid like quasiparticles while the other
band, e.g. related to localized d electrons, undergoes a Mott
transition to an insulating state due to strong electronic
correlations. Such a transition is analogous to the
compressibility-incompressibilty transition of the Cu3d orbital in our
approach.

Because particle density fluctuations scale with the compressibility of
the system the sum rules introduce varying fluctuations of the particle
number in the different metallic states of the HTSC according to our
modeling. The tuning parameter for these fluctuations of the particle
number is doping and the pseudogap state may be considered as a state
with reduced particle density fluctuations as compared with the optimal
and overdoped state. Moreover, the partial incompressibility of the
Cu$3d$ orbitals competes with superconductivity because the phase
fluctuations $\Delta\phi$ of the order parameter grow with decreasing
particle fluctuations $\Delta N$ due to the uncertainty relation
$\Delta N\,\Delta\phi\sim 1$.

Qualitatively, the incompressibility of the Cu$3d$ states also
corresponds to a compartmentalisation of configuration space. This
means that some parts of direct space cannot be approached or are
hardly accessible to the charge carriers. On the other hand, this
should be accompanied by a reciprocal compartmentalisation of
momentum space, according to the uncertainty relationship $\Delta
x\,\Delta k\sim 1$. As a consequence the momentum should have
corresponding fluctuations and the FS of the conventional metallic
state becomes fuzzy when passing to the underdoped state and
possibly reconstructs. ARPES studies and quantum-oscillation
measurements seem to reflect this in form of a reconstruction of
the large FS in the optimal and overdoped state into small Fermi
surface pockets with Fermi-liquid like charge carriers in the
underdoped pseudogap state characterized by a loss of density of
states at the Fermi level consistent with our modeling.

In our orbital based local picture the pseudogap state of the
cuprates looks like a "two component" electronic structure where a
real space organisation of the low lying charge excitations is
achieved in form of a metallic charge response by mobile holes on
the oxygen network in the CuO plane that is blocked at the
incompressible insulator-like Cu sites. Inspection of the sum rule
in equation (\ref{5}) shows that the incompressible Cu$3d$ states
and the compressible O$2p$ states are entangled by the
off-diagonal elements of the polarizability matrix. So our
modelling of the pseudogap state is consistent with a strange not
overall compressible metallic state coexisting with atom-like,
incompressible Cu$3d$ orbitals similar as in the undoped,
insulating state favoring short ranged magnetic spin correlations
and corresponding excitations, typically spin fluctuations. The
latter should be damped by electron-hole charge excitations of the
compressible metallic O$2p$ states.

In $k$-space language such a strange mixed locally
compressible-incompressible state should result in a heavily
renormalized electronic bandstructure and Fermi surface (FS) as
compared with the more conventional overall compressible Fermi liquid
state with a large FS at higher doping. The pseudogap state displaying
properties characteristic of both, insulators and correlated metals
evolves from the incompressible insulating state with no low energy
charge excitations and seems to promote the tendency to charge and spin
ordering that breaks rotational and translational symmetry, e.g. in
form of stripes. The latter appear to be quite common in underdoped
cuprates. In this context it is interesting to point out that the
stripe-like modulation observed in \cite{Kohsaka07} is strongest on the
oxygen orbitals. So, the latter are very essential to understand the
physics of charge order in the cuprates and also the charge response as
shown below. Finally, the enhanced phase fluctuations due to the
incompressibility of the Cu$3d$ orbital favour pairing correlations in
the normal state, possibly incoherent preformed pairs as discussed in
the literature.

\begin{figure}
 \includegraphics[]{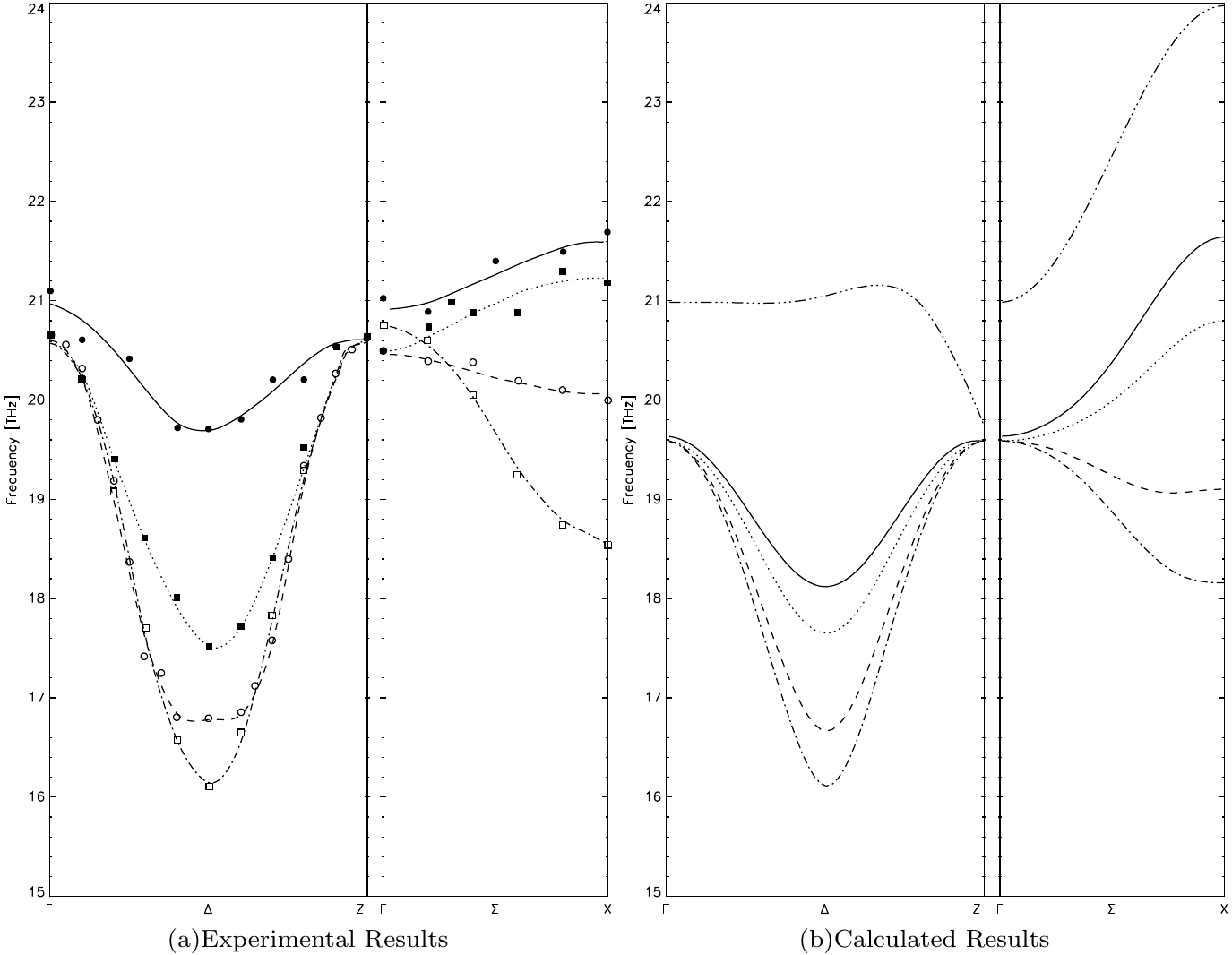}%
\caption{(a) Experimental results for the highest $\Delta_{1}$ and
$\Sigma_{1}$ branch of La$_{2-x}$Sr$_{x}$CuO$_{4}$ for various doping
levels (Ref. 4). $\bullet$: $x\,=\,0.00$, $\blacksquare$: $x\,=\,0.10$,
$\circ$: $x\,=\,0.15$, $\Box$: $x\,=\,0.30$. The lines are a guide to
the eye. (b) Calculated results of the phonon branches shown in (a) as
obtained with our model approach for the electronic state of the HTSC's
\cite{Falter06}. $-\!\!\!-\!\!\!-$: insulating state,
$\cdot\!\cdot\!\cdot$: underdoped state, $-\!-\!-$: optimally doped
state, $-\cdot-$: overdoped state. For comparison, the results for the
RIM are shown $(-\cdot\!\cdot\!\cdot)$ in order to demonstrate the
crucial influence of the {\em nonlocal} EPI effects in form of CF's on
the dispersion.}\label{fig01}
\end{figure}

Our modeling of the density response is strongly supported by
corresponding calculations of phonon dynamics
\cite{Falter06,Bauer08}, which compare well with the experimental
data. This is shown for $p$-doped LaCuO in case of the strongly
doping dependent OBSM phonon anomalies, see \fref{fig01}, where
the experimental results for the highest $\Delta_{1}$ and
$\Sigma_{1}$ branches of La$_{2-x}$Sr$_{x}$CuO$_{4}$ are shown for
various doping levels \cite{Pint05} and compared with the
calculated results. The high-frequency OBSM are generally known to
display an anomalous softening upon doping in the cuprates
\cite{Uchi4,Graf08,Pint98,Pint99,Reich96,Fukuda05,Pint06,McQueeney01,d'Astuto02,d'Astuto03,Braden05,Bauer10}.
Such a behaviour of the experimental phonon dispersion is clearly
visible for LaCuO in \fref{fig01}a and is very well accounted for
by the calculation in our model for the electronic state in terms
of incompressibility-compressibility transitions, see
\fref{fig01}b. We also have included in \fref{fig01}b the phonon
dispersion of a rigid-ion model (RIM) for LaCuO.

It seems appropriate to explain at least the ideas of our modeling
of phonon dynamics and charge response of the cuprates. A
quantitative treatment of the theory and the modeling would be too
extensive and can be found in Refs. \cite{Falter06,Bauer08}.

The rigid part of the electronic charge response and the EPI is
approximated by an ab initio rigid-ion model (RIM), taking into
account covalent ion softening in terms of (static) effective
ionic charges calculated from a tight-binding analysis. In
addition, scaling of the short range part of certain pair
potentials between the ions is performed to simulate further
covalence effects in the calculation in such a way that the
energy-minimized structure is as close as possible to the
experimental one. The RIM with the corrections just mentioned then
serves as an unbiased reference system for the description of the
cuprate superconductors and can be considered as a first
approximation for the insulating state of these compounds because
of the strong ionic nature of bonding in the cuprates. Starting
with such a rigid reference system nonlocal, electronic
polarisation processes are introduced in the form of more or less
localized electronic charge fluctuations (CF's) at the outer
shells of the ions. Especially in the metallic state of the
cuprates the latter dominate the nonlocal contribution of the
charge response and the EPI and are particularly important in the
CuO planes. In addition, anisotropic dipole fluctuations are
admitted in our approach. They prove to be specifically of
interest for the ions in the ionic layers.

Comparing in \fref{fig01}b the results of the phonon dispersion of
the RIM, where only local EPI effects are considered, with the
full calculation we can extract the crucial influence of the
nonlocal EPI effects generated by the charge fluctuations. For
example we find a very strong softening of the $\Delta_{1}/2$
bond-stretching mode (half-breathing mode) increasing with doping
up to about 5 THz. Comparing the calculated results with the
experiments in \fref{fig01}a we find a good agreement. In order to
achieve such a realistic description of the phonon modes, both,
nonlocal EPI effects and multi-orbital effects are required. The
comparison with the results from the RIM demonstrates that
nonlocality is needed and from our detailed calculations we find
that the CF's of the more delocalized O$2p$ and Cu$4s$ orbital
contribute significantly to the softening of the modes, besides
the more localized Cu$3d$ orbital. The strong softening of the
phonon anomalies is accompanied by corresponding dynamic charge
inhomogenities in form of charge stripes  along the CuO bonds. The
modes are not really soft but one may consider these dynamic
stripes as precursors of static charge stripe order with ordering
vectors corresponding to the wavevectors of the phonon anomalies.

\begin{figure}
 \includegraphics[]{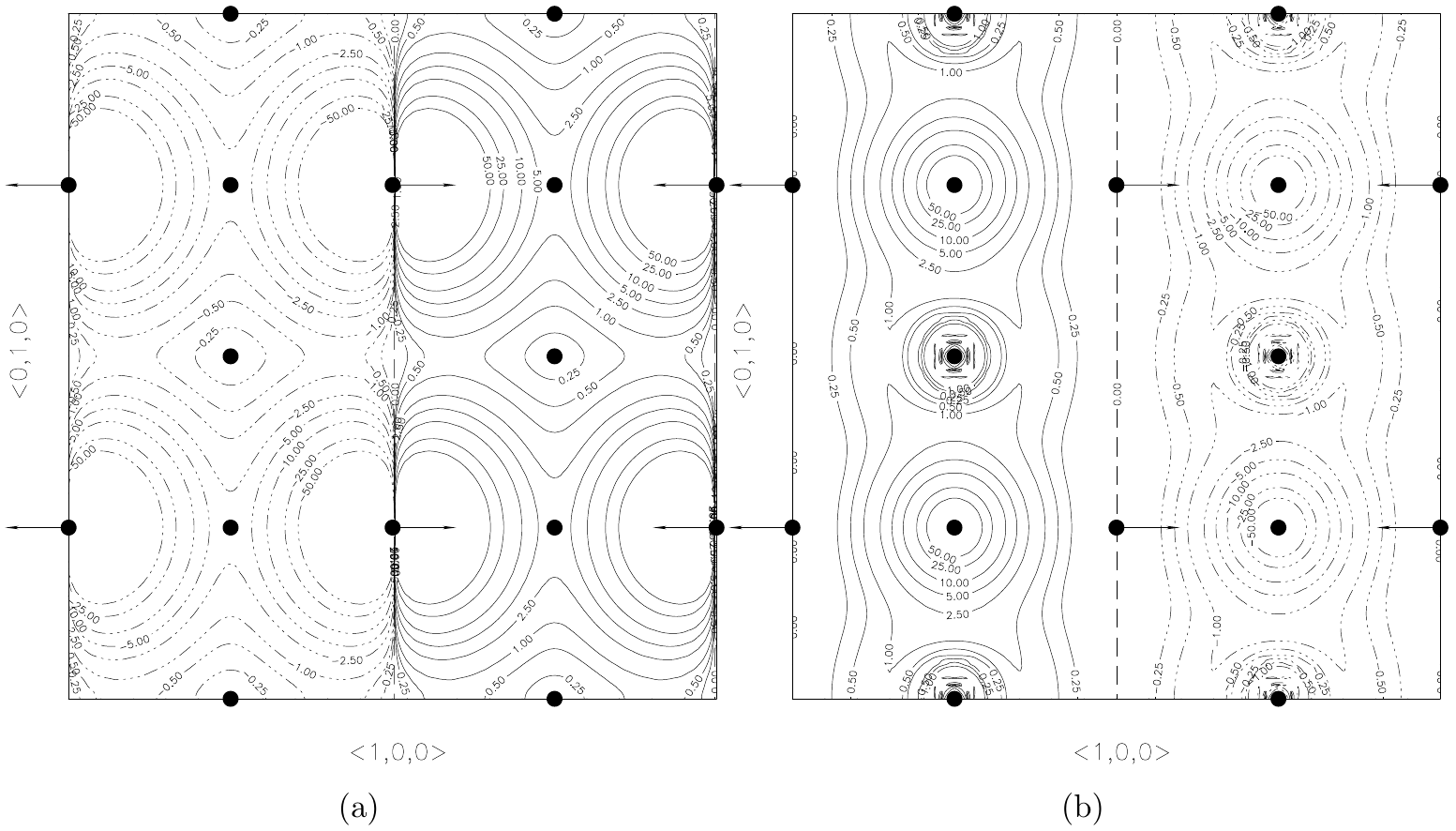}%
\caption{(a) Contour plot in the CuO plane of the displacement induced
charge density redistribution $\delta\rho$ for the $\Delta_{1}/2$ mode
within the rigid ion model (RIM) for La$_{2}$CuO$_{4}$. (b) Same as
(a), but for the nonlocal part of the charge response in the optimally
doped state due to nonlocal EPI effects in the half-breathing mode
$\Delta_{1}/2$. The displaced ions are indicated by a dot with an
arrow. Full lines ($-\!\!\!-\!\!\!-)$ mean that electrons are
accumulated in the corresponding region of space and the broken dotted
lines ($-\!\cdot\!\cdot\!\cdot\!-$) indicate regions where the
electrons are pushed away. $\delta\rho$ is given in units of
$10^{-4}\,e^{2}/a_{B}^{3}$.}\label{fig02}
\end{figure}

Finally, it should be remarked that a constructive interplay of
electron-ion interaction with short-ranged antiferromagnetic order
is important for the Cu sites in the underdoped and insulating
state, because electrons are transferred from the Cu ion where the
Cu bond is compressed to the Cu ion where the bond is stretched,
see e.g. \fref{fig02}. Both ions have opposite spin in the
antiferromagnetic case. In contrast if a ferromagnetic order would
be present the Pauli blocking would suppress the transport of
electrons with the same spin and consequently also the charge
fluctuations and the corresponding softening of the phonon modes.

Modelling of the electronic polarizability
$\Pi_{\kappa\kappa'}\,(\vc{q}\,)$ for the insulating and pseudogap
state is needed because a quantitative calculation of
$\Pi_{\kappa\kappa'}\,(\vc{q}\,)$ is not available due to the
missing of a rigorous knowledge of this state. On the other hand,
for the more conventional metallic state
$\Pi_{\kappa\kappa'}\,(\vc{q}\,)$  can be calculated within the
local density approximation (LDA), however, with some important
corrections concerning the electronic $c$-axis response which is
overestimated in typical LDA-based calculations \cite{Bauer09}.
Concerning this case it should be remarked that a comparison of
the calculated OBSM anomalies from the model in \fref{fig01}b with
corresponding results obtained within the framework of the
corrected LDA \cite{Bauer09}, taking into account the full
$\vc{q}$ dependence of the polarizability matrix, leads to a good
agreement. This also speaks in favour of our modeling of the
charge response in terms of the compressibility sum rules, in
which only the longwavelength limit of the polarizability is
considered.

In \fref{fig02} we display the calculated phonon induced charge
redistribution of the $\Delta_{1}/2$ bond-stretching mode for the
optimal doped state. Figure \ref{fig02}a illustrates the effect of
the {\em local} EPI within the RIM connected with the displacement
of the $O_{x}$ ions in the CuO plane, i.e. a stretching or
compression of the CuO bond, respectively, (half-breathing mode
(HBM)). The local EPI reveals a charge redistribution of dipole
character  localized only at the ions as can be expected from the
rigid shift of the ionic densities in the RIM during the phonon
mode.


From \fref{fig02}a we also obtain the message that no dynamical
charge stripes are excited by local EPI. On the other hand,
\fref{fig02}b that displays the nonlocal part of the charge
response due to nonlocal EPI effects of charge fluctuation type
demonstrates, how the strong nonlocal EPI generates dynamic charge
stripes in the half-breathing mode. Thus, the moving $O_{x}$ ions
in the HBM which are linked up with strong CF's on the silent Cu
and $O_{y}$ ions induce dynamical stripes in the $y$ direction
with a recurrence periond of two lattice constants along the
orthogonal $x$-direction. Note, that there are no changes of the
transfer integral between $d$ and $p$ orbitals for the silent Cu
and $O_{y}$ ions, nevertheless there is a charge transfer.

In case of the full breathing mode $O^{X}_{B}$, at the the $X$ point of
the BZ where the $O_{x}$ and $O_{y}$ ions vibrate in phase along the
bonds against the silent Cu ion \cite{Falter06} nonlocal EPI generates
CF's at the Cu and we obtain an electronic charge transfer from that Cu
ion where the bonds are compressed to the Cu where the bonds are
stretched \cite{Falter05}. While in case of the HBM the charge stripes
point along the $x$- or $y$-direction, respectively, for $O^{X}_{B}$
the stripes point along the diagonals in the CuO plane.

The self-consistent changes of the potential felt by an electron due to
nonlocal EPI also have been calculated for the OBSM and yield a strong
coupling of the order 100 meV \cite{Bauer09}. It should be remarked
that such a strong nonlocal EPI associated with the OBSM, which can
reliably be calculated in the {\it adiabatic} approximation, is further
enhanced by an order of magnitude because of the poorly screened long
range Coulomb interaction and the slow charge dynamics for the polar
modes from the {\it nonadiabatic} sector of phase space around the
$c$-axis where phonon-plasmon mixing is relevant \cite{Bauer09}.

\begin{figure}
 \includegraphics[]{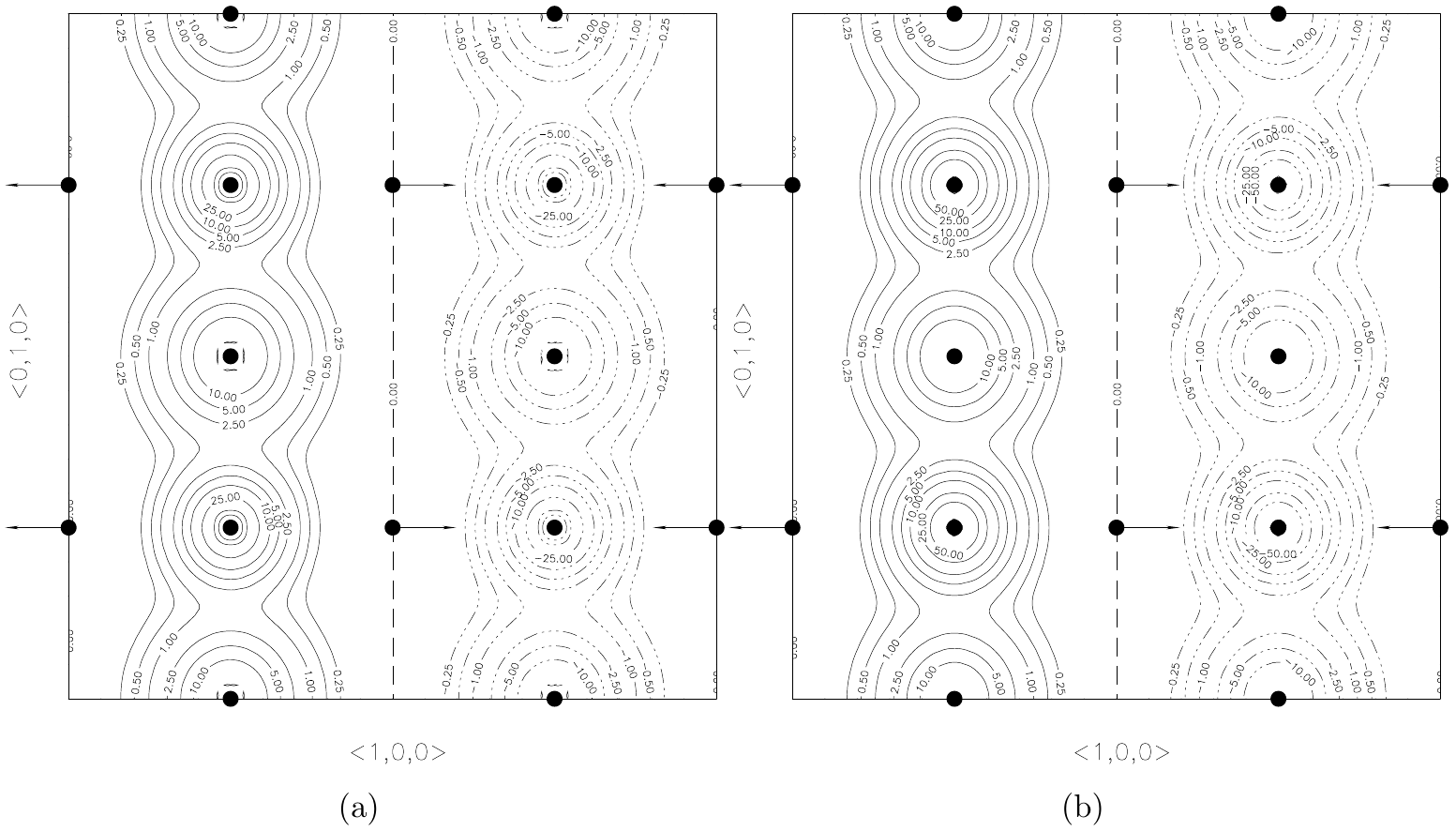}%
\caption{(a) Same as figure 2 but for the nonlocal part of the
charge response $\delta\rho$ for $\Delta_{1}/2$ in the insulating
state ($x=0$). (b) Same as figure 2 but for the nonlocal part of
the charge response $\delta\rho$ for $\Delta_{1}/2$ in the
pseudogap state ($x=0.1$).}\label{fig03}
\end{figure}

In \fref{fig03}a and \fref{fig03}b we present the calculated
results for the nonlocally induced charge redistribution of the
HBM for the insulating and the pseudogap state, which should be
compared to the corresponding results for the optimal doped state
in \fref{fig02}b. First of all we find for {\it all} electronic
states dynamical stripes along the bond direction excited by the
nonlocal  EPI being simultaneously responsible for the
corresponding phonon softening in \fref{fig01}. This is quite
interesting because dynamic charge inhomogenities of this type may
be considered as precursors of static charge stripe order as
recently observed in La$_{2-x}$ Ba$_{x}$ CuO$_{4}$, in a broad
range of doping around $x\,=\,1/8$ with $0.095 \leq x \leq 0.155$
\cite{Huecker11} or in YBa$_2$Cu$_3$O$_y$, respectively,
\cite{Taowu11}.

To the best of our knowledge in overdoped samples static stripe
order has not been observed so far. This would mean that the
pseudogap in the underdoped state may be required to nucleate
dynamic stripes. In our approach this implies that for the
emergence of static stripe order the incompressibility of the
Cu$3d$ orbitals seems to be essential. The latter brings about a
separation of mobile holes populating the compressible O$2p$
states and regions with local Mott- and spin correlations related
to the insulator like Cu$3d$ states. Such stripes producing the
onset of a new periodicity that breaks translational and
rotational symmetry also may help to trigger a reconstruction of
the large hole Fermi surface (FS) into small pockets as detected
by quantum oscillation measurements in underdoped YBa$_{2}$
Cu$_{3}$ O$_{6+x}$, see e.g. \cite{LeBoeuf11} and references
therein. Also from an energetic point of view our modeling of the
pseudogap state is susceptible for the tendency to establish
broken symmetry states with hole-rich compressible and hole-poor
incompressible regions. In general terms lowering of the kinetic
energy occurs in the compressible metallic regions and lowering of
the interaction energy is possible by antiferromagnetic spin
correlations and the Coulomb repulsion between the electrons in
the region of the incompressible insulator-like Cu$3d$ states.
Altogether, in the pseudogap state the doped holes and the
superexchange-coupled spins arrange themselves in a cooperative
way to promote both charge mobility and locally antiferromagnetic
correlations.

Comparing the stripe patterns of the insulating and underdoped state
with that of the more conventional metallic optimal and overdoped state
(not shown) with a large FS as the locus of the crucial low energy
excitations, we find a relatively small contribution of the O$2p$
orbitals degrees of freedom to the density redistribution in the latter
case. The weight of the O$2p$ contribution is significantly enhanced in
the insulating and pseudogap state, see \fref{fig03}a and \fref{fig03}b
reflecting in particular the importance of the O$2p$ orbital in the
pseudogap state. Holes in the pseudogap state primarily populate the
O$2p$ orbitals, while the holes in the more conventional metallic state
gain more and more Cu$3d$ character by an increased hybridization as
can be seen from a large PDOS for Cu$3d$ leading to a dominating
contribution of the charge redistribution at the Cu ion where the
charge response is now metallic, too. See also \cite{Sakurai11} where
it is investigated how the nature of the hole state evolves with
doping. Overall, the stripe pattern is more delocalized in the metallic
state and localization increases towards the insulating state via the
pseudogap state.

In the overdoped state a growing importance of the Cu$4s$ orbital and
accordingly an increase of the Cu$4s$ element of the polarizability
matrix leads to an enhanced softening of the anomalous OBSM, see
\fref{fig01}. The corresponding calculated dynamical stripe pattern is
very similar to that of the optimally doped state in \fref{fig02}b but
somewhat more delocalized because of the increased contribution of the
more delocalized Cu$4s$ contribution.

From our calculations it becomes obvious that a multi-orbital
approach is needed that includes besides Cu$3d$ at least O$2p$
orbitals and for the optimally doped and overdoped state in
addition Cu$4s$ orbitals in order to give a reliable
representation of phonon dynamics and charge response in the
cuprates. Single-band models for the CuO plane believed by many
people to describe the cuprates are not sufficient, in particular
in context with the discussion of stripes. Altogether, in the
optimal and overdoped state increasing of doping enhances the
hybridization between Cu$3d$, Cu$4s$ and O$2p$ orbitals,
respectively, with a dominant weight for Cu$3d$. This state
becomes compressible metallic with a large PDOS at
$\varepsilon_{F}$. So, the particle number fluctuations that
characterize the different competing ground states in our model
are enhanced as compared to the ground state in the pseudogap
regime. Here the hybridization is weaker due to the incompressible
insulator-like Cu$3d$ orbital with a vanishing PDOS at
$\varepsilon_{F}$.

\begin{figure}
 \includegraphics[angle=90,width=\linewidth]{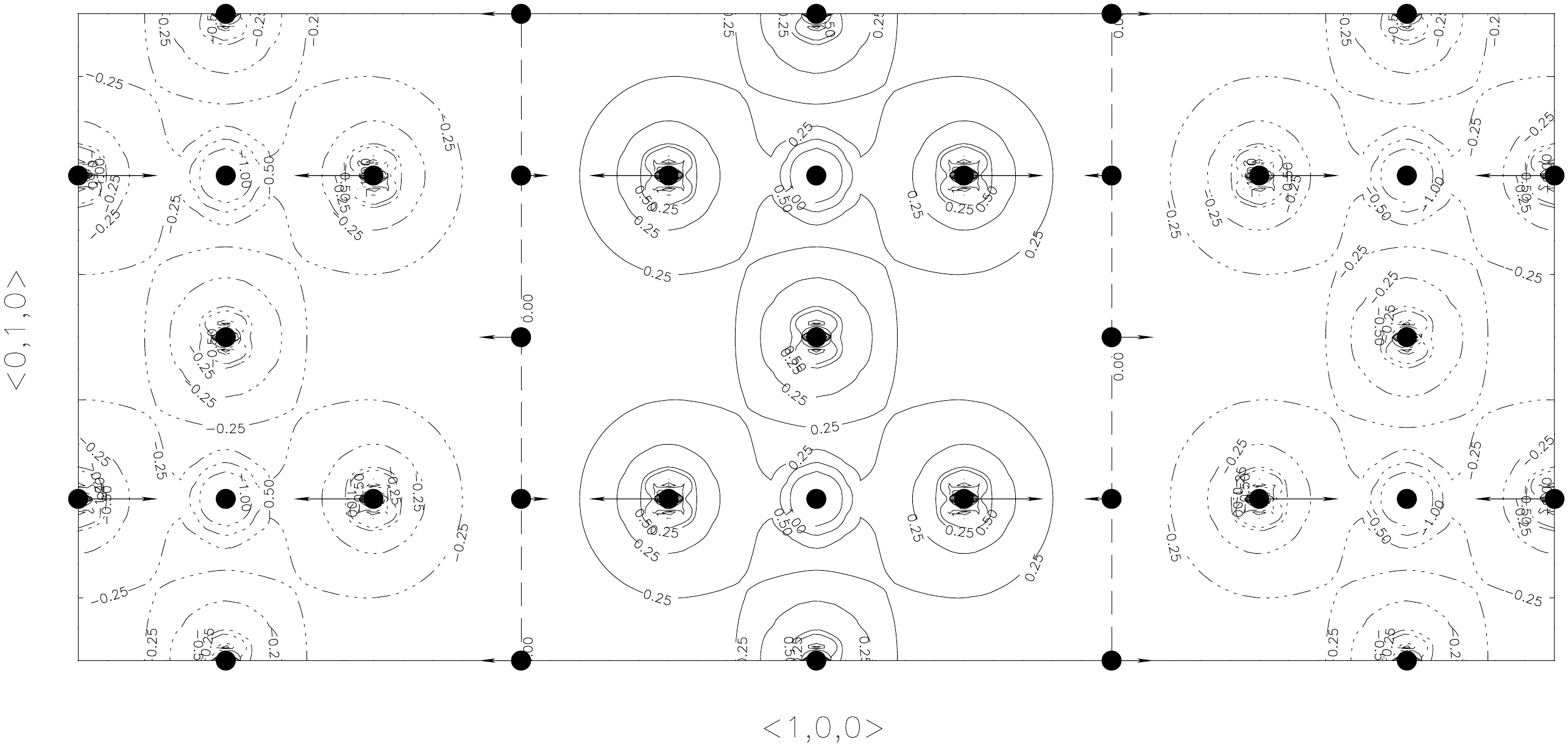}%
\caption{Same as figure 2 but for the nonlocal part of the charge
response $\delta\rho$ for $\Delta_{1}/4$ in the optimally doped
state.}\label{fig04}
\end{figure}

Translational symmetry breaking models of the electronic structure
of the CuO plane generated by charge density wave order of a
certain wavevector are discussed in the literature and used to
characterize Fermi surface reconstructions of a large FS into
small pockets, for recent work see \cite{Harrison11,Yao11}. From
scanning tunneling microscopy (STM), neutron diffraction
experiments and nuclear magnetic resonance measurements the
ordering in many cases occurs with a period of about 4 lattice
constants \cite{Yao11}. This corresponds to the wavevector
$\frac{\pi}{2\,a}\,(1,0,0)$ of the $\Delta_{1}/4$ mode. Therefore
it is meaningful to calculate the nonlocally induced charge
redistribution also for this mode. The result for the optimal
doped state is shown in \fref{fig04}.

We again obtain a stripe like charge pattern with translational
symmetry in the $(0,1,0)$ direction but a four unit cell repeat
distance along the orthogonal direction (1,0,0). Contrasting this
result with the charge stripe pattern as found for the HBM in
\fref{fig02}b the stripes for $\Delta_{1}/4$ are expanded over the
whole unit cell, because the CF's do not vanish at the sites of the
moving oxygens, unlike to $\Delta_{1}/2$. However, the amplitudes of
the CF's are considerably smaller. According to our calculation of the
eigenvectors this is caused by the strongly reduced bond-stretching
amplitude of the oxygen ions in the plane for $\Delta_{1}/4$, while the
displacement of the apex oxygen ions is strongly enhanced in the
$\Delta_{1}/4$ mode as compared with the HBM.

In summary, we have applied our modeling of the doping dependent
electronic states of the cuprates to the calculation of the phonon
induced charge inhomogeneities nonlocally induced by the $\Delta_{1}/2$
and $\Delta_{1}/4$ mode in LaCuO. The calculations have been performed
for the insulating-, pseudogap and the more conventional metallic
state. The compressibility is used as a primary term to characterize
the competing ground states controlled by doping. The strong nonlocal
EPI of CF type excites in all cases investigated dynamical charge
stripes along the CuO bonds. Differences in orbital character of the
charge response in the various states are discussed at hand of the
stripe patterns. Experimentally, charge stripe order has been observed
in a broad range of doping in La$_{2-x}$Ba$_{x}$CuO$_{4}$ and other
cuprates which supports the point of view that dynamical charge stripes
generated by certain strongly coupling OBSM may be considered as
precursors of static stripe order at least outside the overdoped
regime. Such a charge order breaks translational symmetry and may
enforce a reconstruction of a larger hole FS into small pockets.

The interplay between the reduced particle density fluctuations related
to the insulator-like behaviour of the localized atomic-like Cu$3d$
orbital that remains incompressible for low enough $p$-doping and on
the other hand the compressible, metallic O$2p$ orbitals appear to play
an important role for the pseudogap state and the tendency to charge
ordering in the cuprates.


\section*{References}
\begin{thebibliography}{99}
\bibitem{Falter06} Falter C, Bauer T and Schnetg\"oke F, 2006, Phys. Rev. B {\bf 73} 224502
\bibitem{Bauer08} Bauer T and Falter C, 2008, Phys. Rev. B {\bf 77} 144503
\bibitem{Svane92} Svane A 1992 {\em Phys. Rev. Lett} {\bf 68} 1900
\bibitem{Vojta} Vojta M, arXiv: 1202.1913
\bibitem{Kohsaka07} Kohsaka Y {\em et al} 2007 Science {\bf 315} 1380
\bibitem{Pint05} Pintschovius L 2005 {\em Phys. Status Solidi b} {\bf 242} 30
\bibitem{Uchi4} Uchiyama H et al 2004 {\em Phys. Rev. Lett} {\bf 92} 197005
\bibitem{Graf08} Graf J {\em et al} 2008 {\em Phys. Rev. Lett} {\bf 100} 227002
\bibitem{Pint98} Pintschovius L and Reichhardt W 1998 {\em Neutron Scattering in
                 Layered Copper-Oxide Superconductors (Physics and Chemistry of
                 Materials with Low Dimensional Structures vol 20)}
                 ed A Furrer (Dordrecht: Kluwer-Academic)
\bibitem{Pint99} Pintschovius L and Braden M 1999 {\em Phys. Rev. B} {\bf 60} R15039
\bibitem{Reich96} Reichardt W 1996 {\em J. Low Temp. Phys.} {\bf 105} 807
\bibitem{Fukuda05} Fukuda T {\em et al} 2005 {\em Phys. Rev. B} {\bf 71} R060501
\bibitem{Pint06} Pintschovius L {\em et al} 2006 {\em Phys. Rev. B} {\bf 74} 174514
\bibitem{McQueeney01} McQueeney R J {\em et al} 2001 {\em Phys. Rev. Lett.} {\bf 87} 077001
\bibitem{d'Astuto02} d'Astuto {\em et al} 2002 {\em Phys. Rev. Lett.} {\bf 88} 167002
\bibitem{d'Astuto03} d'Astuto {\em et al} 2003 {\em Int. J. Mod. Phys. B} {\bf 17} 484
\bibitem{Braden05} Braden M {\em et al} 2005 {\em Phys. Rev. B} {\bf 72} 184517
\bibitem{Bauer10} Bauer T and Falter C 2010 {\em J. Phys.: Condens. Matter} {\bf 22} 142201
\bibitem{Bauer09} Bauer T and Falter C 2009 {\em Phys. Rev. B} {\bf 80} 094525
\bibitem{Falter05} Falter C 2005 {\em Phys. Stat. Sol (b)} {\bf 242} 78
\bibitem{Huecker11} H\"ucker M {\em et al} 2011 {\em Phys. Rev. B} {\bf 83} 104506
\bibitem{Taowu11} Tao Wu {\em et al} 2011 {\em Nature} {\bf 477} 191
\bibitem{LeBoeuf11} Le Boeuf D {\em et al} 2011 {\em Phys. Rev B} {\bf 83} 054506
\bibitem{Sakurai11} Sakurai Y {\em et al} 2011 Science {\bf 323} 698
\bibitem{Harrison11} Harrison N and Sebastian S E 2011 {\em Phys. Rev. Lett.} {\bf 106} 226402
\bibitem{Yao11} Yao H, Lee D-H and Kivelson S 2011 Phys. Rev. B {\bf 84} 012507
\end{thebibliography}
\end{document}